\documentclass[article]{jss}

\usepackage{graphicx}
\usepackage{amsfonts}
\usepackage{amssymb}
\usepackage{amsmath}
\usepackage{algorithm}
\usepackage{multirow}

\newcommand{\given}{\mid}
\newcommand{\Pa}{\boldsymbol\Pi}
\newcommand{\B}{\mathcal{B}}
\newcommand{\N}{\mathcal{N}}
\newcommand{\V}{\mathcal{V}}
\newcommand{\COR}{\mathsf{COR}}

\author{Marco Scutari\\University of Oxford}
\title{Bayesian Network Constraint-Based Structure Learning Algorithms: 
  Parallel and Optimised Implementations in the \pkg{bnlearn} \proglang{R}
  Package}

\Plainauthor{Marco Scutari}
\Plaintitle{Bayesian Network Constraint-Based Structure Learning Algorithms:
  Parallel and Optimised Implementations in the bnlearn R Package}
\Shorttitle{Bayesian Network Learning: Parallel and Optimised Implementations}

\Abstract{
It is well known in the literature that the problem of learning the structure
of Bayesian networks is very hard to tackle: its computational complexity is
super-exponential in the number of nodes in the worst case and polynomial in
most real-world scenarios.

Efficient implementations of score-based structure learning benefit from past
and current research in optimisation theory, which can be adapted to the task
by using the network score as the objective function to maximise. This is not
true for approaches based on conditional independence tests, called 
constraint-based learning algorithms. The only optimisation in widespread use,
backtracking, leverages the symmetries implied by the definitions of 
neighbourhood and Markov blanket.

In this paper we illustrate how backtracking is implemented in recent versions
of the \pkg{bnlearn} \proglang{R} package, and how it degrades the stability
of Bayesian network structure learning for little gain in terms of speed. As an
alternative, we describe a software architecture and framework that can be used
to parallelise constraint-based structure learning algorithms (also implemented
in \pkg{bnlearn}) and we demonstrate its performance using four reference 
networks and two real-world data sets from genetics and systems biology. We
show that on modern multi-core or multiprocessor hardware parallel 
implementations are preferable over backtracking, which was developed when
single-processor machines were the norm.
}
\Keywords{Bayesian networks, structure learning, parallel programming, \proglang{R}}
\Plainkeywords{Bayesian networks, structure learning, parallel programming, R}

\Address{
  Marco Scutari\\
  Department of Statistics\\
  University of Oxford\\
  1 South Parks Road \\
  OX1 3TG Oxford, United Kingdom\\
  E-mail: \email{scutari@stats.ox.ac.uk}\\
  URL: \url{http://www.bnlearn.com/about}
}

\begin{document}

\section{Background and notations}

Bayesian networks (BNs) are a class of \textit{graphical models} 
\citep{pearl,koller} composed by a set of random variables
$\mathbf{X} = \{X_i, i = 1, \ldots, m\}$ and a \textit{directed acyclic graph}
(DAG), denoted $G = (\mathbf{V}, A)$ where $\mathbf{V}$ is the \textit{node set}
and $A$ is the \textit{arc set}. The probability distribution of $\mathbf{X}$
is called the \textit{global distribution} of the data, while those associated
with individual $X_i$s are called \textit{local distributions}. Each node in 
$\mathbf{V}$ is associated with one variable, and they are referred to
interchangeably. The directed arcs in $A$ that connect them are denoted as
``$\rightarrow$'' and represent direct stochastic dependencies; so if there is
no arc connecting two nodes, the corresponding variables are either marginally
independent or conditionally independent given (a subset of) the rest. As a 
result, each local distribution depends only on a single node $X_i$ and on its
parents (i.e., the nodes $X_j, j \neq i$ such that $X_j \rightarrow X_i$, here
denoted $\Pa_{X_i}$):
\begin{equation}
  \label{eq:local}
  \Prob\left(\mathbf{X}\right) = \prod_{i=1}^m \Prob\left(X_i \given \Pa_{X_i}\right).
\end{equation}
Common choices for the local and global distributions are multinomial variables
\citep[discrete BNs,][]{heckerman}; univariate and multivariate normal variables
\citep[Gaussian BNs,][]{geiger}; and, less frequently, a combination of the two 
\citep[conditional Gaussian (CG) BNs,][]{cgbn}. In the first case, the parameters
of interest are the conditional probabilities associated with each variable,
represented as \textit{conditional probability tables} (CPTs); in the second,
the parameters of interest are the \textit{partial correlation coefficients}
between each variable and its parents. As for CG BNs, the parameters of interest
are again partial correlation coefficients, computed for each node conditional
on its continuous parents for each configuration of the discrete parents.

The key advantage of the decomposition in Equation (\ref{eq:local}) is to make
\textit{local computations} possible for most tasks, using just a few variables
at a time regardless of the magnitude of $m$. A related quantity that works to
the same effect is the \textit{Markov blanket} of each node $X_i$, defined as
the set $\B(X_i)$ of nodes which graphically separates $X_i$ from all other
nodes $\mathbf{V} \setminus \{X_i, \B(X_i)\}$ \citep[][p. 97]{pearl}. In BNs
such a set is uniquely identified by the parents ($\Pa_{X_i}$), the children
(i.e., the nodes $X_j, j \neq i$ such that $X_i \rightarrow X_j$) and the
spouses of $X_i$ (i.e., the nodes that share a child with $X_i$). By definition,
$X_i$ is independent of all other nodes given $\B(X_i)$, thus making them
redundant for inference on $X_i$.

The task of fitting a BN is called \textit{learning} and is generally
implemented in two steps.

The first is called \textit{structure learning}, and consists in finding the DAG
that encodes the conditional independencies present in the data. This has been
achieved in the literature with \textit{constraint-based}, \textit{score-based}
and \textit{hybrid} algorithms; for an overview see \citet{koller} and 
\citet{crc13}. Constraint-based algorithms are based on the seminal work of
Pearl on causal graphical models and his Inductive Causation algorithm
\citep[IC, ][]{ic}, which provides a framework for learning the DAG of a BN using 
conditional independence tests under the assumption that graphical separation
and probabilistic independence imply each other (the \textit{faithfulness}
assumption). Tests in common use are the mutual information test (for discrete
BNs) and the exact Student's $t$ test for correlation (for GBNs). Score-based
algorithms represent an application of heuristic optimisation techniques: each
candidate DAG is assigned a network score reflecting its goodness of fit, which
the algorithm then attempts to maximise. BIC and posterior probabilities arising
from various priors are typical choices. Hybrid algorithms use both conditional
independence tests and network scores, the former to reduce the space of
candidate DAGs and the latter to identify the optimal DAG among them. Some
examples are PC \citep[named after its inventors Peter Spirtes and Clark 
Glymour;][]{spirtes}, Grow-Shrink \cite[GS;][]{gs}, Incremental Association 
\citep[IAMB;][]{iamb}, Interleaved IAMB \citep[Inter-IAMB;][]{fastiamb2}, 
hill-climbing and tabu search \citep{norvig}, Max-Min Parents \& Children
\citep[MMPC;][]{mmhc} and the Semi-Interleaved HITON-PC \citep[SI-HITON-PC,
from the Greek ``hiton'' for ``blanket'';][]{hitonpc}.
These algorithms and more are implemented across several \proglang{R} packages,
such as \pkg{bnlearn} \citep[all of the above except PC;][]{jss09},
\pkg{deal} \citep[hill-climbing;][]{deal}, \pkg{catnet} and \pkg{mugnet}
\citep[simulated annealing;][]{catnet,mugnet}, \pkg{pcalg}
\citep[PC and causal graphical model learning algorithms;][]{pcalg} and
\pkg{abn} \citep[hill climbing and exact algorithms;][]{abn}. Further extensions
to model dynamic data are implemented in \pkg{ebdbNet} \citep{ebdbNet},
\pkg{G1DBN} \citep{G1DBN} and \pkg{ARTIVA} \citep{artiva} among others.

The second step is called \textit{parameter learning} and, as the name suggests,
deals with the estimation of the parameters of the global distribution. Since
the graph structure is known from the previous step, this can be done efficiently
by estimating the parameters of the local distributions. With the exception of
\pkg{bnlearn}, which has a separate \code{bn.fit} function, \proglang{R} packages
automatically execute this step along with structure learning.

Most problems in BN theory have a computational complexity that, in the worst
case, scales exponentially with the number of variables. For instance, both
structure learning \citep{nphard,nphard3} and inference \citep{nphard2,nphard4}
are NP-hard and have polynomial complexity even for sparse networks. This is
especially problematic in high-dimensional settings such as genetics and systems
biology, in which BNs are used for the analysis of gene expressions
\citep{friedman} and protein-protein interactions \citep{jansen,sachs}; for
integrating heterogeneous genetic data \citep{integrative}; and to determine
disease susceptibility to anemia \citep{anemia} and hypertension \citep{malovini}.

Even though algorithms in recent literature are designed taking scalability
into account, it is often impractical to learn BNs from data containing more
than few hundreds of variables without restrictive assumptions on either the
structure of the DAG or the nature of the local distributions. Two ways to
address this problem are:
\begin{enumerate}
  \item \textit{optimisations:} reducing the number of conditional independence
    tests and network scores computed from the data, either by skipping
    redundant ones or by limiting local computations to a few variables;
  \item \textit{parallel implementations:} splitting learning across multiple
    cores and processors to make better use of modern multi-core and
    multiprocessor hardware.
\end{enumerate}
As far as score-based learning algorithms are concerned, both possibilities
have been explored and a wide range of solutions proposed, from efficient
caching using decomposable scores \citep{hcopt}, to parallel meta-heuristics
\citep{metaheuristic} and integer programming \citep{cussens}. The same cannot
be said of constraint-based algorithms; even recent ones such as SI-HITON-PC,
while efficient, are still implemented with basic backtracking as the only
optimisation. We will examine them and their implementations in Section 
\ref{sec:backtracking}, arguing that backtracking provides only modest speed
gains and increases the variability of the learned DAGs. As an alternative, in
Section  \ref{sec:parallel} we describe a software architecture and framework
that can be used to create parallel implementations of constraint-based
algorithms that scale well on large data sets and do not suffer from this
problem. In both sections we will focus on the \pkg{bnlearn} package because
it provides the widest choice of algorithms and implementations among those
of interest for this paper.

\section{Constraint-based structure learning and backtracking}
\label{sec:backtracking}

All constraint-based structure learning algorithms share a common three-phase
structure inherited from the IC algorithm through PC and GS; it is illustrated
in Algorithm \ref{algo:ic}. The first, optional, phase consists in learning the
Markov blanket of each node to reduce the number of candidate DAGs early on.
Any algorithm for learning Markov blankets can be plugged in step \ref{pt1}
and extended into a full BN structure learning algorithm, as originally 
suggested in \citet{gs} for GS. Once all Markov blankets have been learned, 
they are checked for consistency (step \ref{pt2}) using their symmetry;
by definition $X_i \in \B(X_j) \Leftrightarrow X_j \in \B(X_i)$. Asymmetries
are corrected by treating them as false positives and removing the offending
nodes from each other's Markov blankets.

The second phase learns the \textit{skeleton} of the DAG, that is, it identifies
which arcs are present in the DAG modulo their direction. This is equivalent to
learning the \textit{neighbours} $\N(X_i)$ of each node: its parents and
children. As illustrated in step \ref{pt3}, the absence of a set of nodes
$\mathbf{S}_{X_i X_j}$ that separates a particular pair $X_i, X_j$ implies
that either $X_i \rightarrow X_j$ or $X_j \rightarrow X_i$. Separating sets are
considered in order of increasing size to keep computations as local as
possible. Furthermore, if $\B(X_i)$ and $\B(X_j)$ are available from steps
\ref{pt1} and \ref{pt2} the search space can be greatly reduced because $\N(X_i)
\subseteq \B(X_i)$. On the one hand, if $X_j \notin \B(X_i)$ by definition $X_i$
is separated from $X_j$ by $\mathbf{S}_{X_i X_j} = \B(X_i)$. On the other hand,
if $X_j \in \B(X_i)$ most candidate sets can be disregarded because we know that
$\mathbf{S}_{X_i X_j} \subseteq \B(X_i) \setminus X_j$ and $\mathbf{S}_{X_i X_j}
\subseteq \B(X_j) \setminus X_i$; both sets are typically much smaller than
$\mathbf{V}$. With the exception of the PC algorithm, which is structured
exactly as described in step \ref{pt3}, constraint-based algorithms learn the 
skeleton by learning each $\N(X_i)$ and then enforcing symmetry (step \ref{pt4}).

Finally, in the third phase arc directions are established as in \citet{meek}.
It is important to note that, for some arcs, both directions are equivalent in
the sense that they identify equivalent decompositions of the global
distribution. Therefore, some arcs will be left undirected and the algorithm
will return a \textit{completed partially directed acyclic graph} identifying
an \textit{equivalence class} containing multiple DAGs. Such a class is uniquely
identified by the skeleton learned in steps \ref{pt3} and \ref{pt4}, and by the
\textit{v-structures} $\V_l$ of the form $X_i \rightarrow X_k \leftarrow X_j$,
$X_i \notin \N(X_j)$ learned in step \ref{pt5} \citep{chickering}. Additional
arc directions are inferred indirectly in step \ref{pt6} by ruling out those
that would introduce additional v-structures (which would have been identified
in step \ref{pt5}) or cycles (which are not allowed in DAGs).

\begin{algorithm}[t!]
\caption{A template for constraint-based structure learning algorithms}
\label{algo:ic}
\normalsize
\vspace{2mm}
\textbf{Input:} a data set containing the variables $X_i, i = 1, \ldots, m$.

\textbf{Output:} a completed partially directed acyclic graph.
\newline

\textbf{Phase 1: learning Markov blankets} (optional).
\begin{enumerate}
  \item For each variable $X_i$, learn its Markov blanket $\B(X_i)$. \label{pt1}
  \item Check whether the Markov blankets $\B(X_i)$ are symmetric, e.g.,
    $X_i \in \B(X_j) \Leftrightarrow X_j \in \B(X_i)$. Assume that nodes for
    whom symmetry does not hold are false positives and drop them from each
    other's Markov blankets. \label{pt2}
\end{enumerate}
\textbf{Phase 2: learning neighbours.}
\begin{enumerate}
  \setcounter{enumi}{2}
  \item For each variable $X_i$, learn the set $\N(X_i)$ of its neighbours
    (i.e., the parents and the children of $X_i$). Equivalently, for each pair 
    $X_i, X_j, i \neq j$ search for a set $\mathbf{S}_{X_i X_j} \subset
    \mathbf{V}$ (including $\mathbf{S}_{X_i X_j} = \varnothing$) such that
    $X_i$ and $X_j$ are independent given $\mathbf{S}_{X_i X_j}$ and $X_i,
    X_j \notin \mathbf{S}_{X_i X_j}$. If there is no such a set, place an
    undirected arc between $X_i$ and $X_j$ ($X_i - X_j$). If $\B(X_i)$ and
    $\B(X_j)$ are available from points \ref{pt1} and \ref{pt2}, the search
    for $\mathbf{S}_{X_i X_j}$ can be limited to the smallest of $\B(X_i)
    \setminus X_j$ and $\B(X_j) \setminus X_i$. \label{pt3}
  \item Check whether the $\N(X_i)$ are symmetric, and correct asymmetries as
    in step \ref{pt2}. \label{pt4}
\end{enumerate}
\textbf{Phase 3: learning arc directions.}
\begin{enumerate}
  \setcounter{enumi}{4}
  \item For each pair of non-adjacent variables $X_i$ and $X_j$ with a common
    neighbour $X_k$, check whether $X_k \in \mathbf{S}_{X_i X_j}$. If not, set
    the direction of the arcs $X_i - X_k$ and $X_k - X_j$ to $X_i \rightarrow X_k$
    and $X_k \leftarrow X_j$ to obtain a v-structure
    $\V_l = \{ X_i \rightarrow X_k \leftarrow X_j \}$. \label{pt5}
  \item Set the direction of arcs that are still undirected by applying
    the following two rules recursively: \label{pt6}
    \begin{enumerate}
      \item If $X_i$ is adjacent to $X_j$ and there is a strictly directed path
        from $X_i$ to $X_j$ (a path leading from $X_i$ to $X_j$ containing no 
        undirected arcs) then set the direction of $X_i - X_j$ to $X_i 
        \rightarrow X_j$.
      \item If $X_i$ and $X_j$ are not adjacent but $X_i \rightarrow X_k$ and
        $X_k - X_j$, then change the latter to $X_k \rightarrow X_j$.
    \end{enumerate}
\end{enumerate}
\end{algorithm}

Even in such a general form, we can see that Algorithm \ref{algo:ic} performs
many checks for graphical separation that are redundant given the symmetry of
the $\B(X_i)$ and the $\N(X_i)$. Intuitively, once we have concluded that
$X_j \notin \B(X_i)$ there is no need to check whether $X_i \in \B(X_j)$ in
step \ref{pt1}; and similar considerations can be made for neighbours in step
\ref{pt3}. In practice, this translates to redundant independence tests
computed on the data. Therefore, up to version 3.4 \pkg{bnlearn} implemented 
\textit{backtracking} by assuming $X_1, \ldots, X_m$ were processed sequentially
and enforcing symmetry by construction (e.g., if $i < j$ then $X_j \notin
\B(X_i) \Rightarrow X_i \notin \B(X_j)$ and $X_j \in \B(X_i) \Rightarrow X_i
\in \B(X_j)$). While this approach on average reduces the number of tests by 
a factor of $2$, it also introduces a false positive or a false negative in
the learning process for every type I or type II error in the independence
tests. As long as BN learning was only feasible for simple data sets (due to
limitations in computational power and algorithm efficiency), and the focus
was on probabilistic modelling, the overall error rate was still acceptable;
but the increasing prevalence of causal modelling on ``small $n$, large $p$''
data sets in many fields requires a better approach. One such is described
in \citet[][p. 46]{mmhc} and implemented in \pkg{bnlearn} from version 3.5.
It modifies steps \ref{pt1} and \ref{pt3} as follows:
\begin{itemize}
  \item If $X_j \notin \B(X_i), i < j$, then do not consider $X_i$ for inclusion
    in $\B(X_j)$; and if $X_j \notin \N(X_i)$, then do not consider $X_i$
    for inclusion in $\N(X_i)$.
  \item If $X_j \in \B(X_i)$ , then consider $X_i$ for inclusion in $\B(X_j)$ by
    initialising $\B(X_j) = \{ X_i \}$; and if $X_j \in \N(X_i)$ then initialise
    $\N(X_j) = \{ X_i \}$. Note that in both cases $X_i$ can be discarded in the
    process of learning $\B(X_j)$ and $\N(X_j)$.
\end{itemize}
Even in this form, backtracking has the undesirable effect of making structure
learning depend on the order the variables are stored in the data set, which
has been shown to increase errors in the PC algorithm \citep{colombo}. In
addition, backtracking provides only a modest speed increase compared to a
parallel implementation of steps \ref{pt1}-\ref{pt4}; we will compare the
respective running times in Section \ref{sec:parallel}. However, it is easy
to implement side-by-side with the original versions of constraint-based
structure learning algorithms. Such algorithms are typically described only
at the node level, that is, they define how each $\B(X_i)$ and $\N(X_i)$ is
learned and then they combine them as described in Algorithm \ref{algo:ic}. 
\pkg{bnlearn} exports two functions that give access to the corresponding
backends: \code{learn.mb} to learn a single $\B(X_i)$ and \code{learn.nbr} to
learn a single $\N(X_i)$. The old approach to backtracking essentially
whitelisted all nodes such that $X_j \in \B(X_i)$ and blacklisted all nodes
such that $X_j \notin \B(X_i)$ for each $X_i$.
\begin{Code}
R> library(bnlearn)
R> data(learning.test)
R> learn.nbr(x = learning.test, method = "si.hiton.pc", node = "D",
+              whitelist = c("A", "C"), blacklist = "B")
\end{Code}
For example, in the code above we learn $\N(\mathtt{D})$ and, assuming we
already learned $\N(\mathtt{A})$, $\N(\mathtt{B})$ and $\N(\mathtt{C})$, we
\code{whitelist} and \code{blacklist} \code{A}, \code{B} and \code{C} depending
on whether \code{D} was one of their neighbours or not. The remaining nodes
in the BN are neither whitelisted nor blacklisted and are then tested for
conditional independence. By contrast, the current approach initialises
$\N(\mathtt{D})$ as $\{\mathtt{A}, \mathtt{C}\}$ but does not whitelist those
nodes.
\begin{Code}
R> learn.nbr(x = learning.test, method = "si.hiton.pc", node = "D",
+              blacklist = "B", start = c("A", "C"))
\end{Code}
As a result, both \code{A} and \code{C} can be removed from $\N(\mathtt{D})$ by
the algorithm. The vanilla, non-optimised equivalent for the same learning
algorithm would be
\begin{Code}
R> learn.nbr(x = learning.test, method = "si.hiton.pc", node = "D")
\end{Code}
which does not include any information from $\N(\mathtt{A})$, $\N(\mathtt{B})$
or $\N(\mathtt{C})$. The syntax for \code{learn.mb} is identical, and will be
omitted for brevity. The only other \proglang{R} package implementing general
constraint-based structure learning, \pkg{pcalg}, implements the PC algorithm
as a monolithic function and does not export the backends which are used to
learn the presence of an arc between a pair of nodes. Furthermore, as we noted
above, PC is implemented differently from other constraint-based algorithms
and is usually modified with different optimisations than backtracking; see,
for instance, the interleaving described in \citet{colombo}.

In the remainder of this section we will focus on the effects of backtracking on
learning the skeleton of the DAG, because steps \ref{pt1}-\ref{pt4} comprise
the vast majority of the overall conditional independence tests and thus control
most of the variability of the DAG. To investigate it, we used \pkg{bnlearn}
and $5$ reference BNs of various size and complexity from
\url{http://www.bnlearn.com/bnrepository}:
\begin{itemize}
  \item ALARM \citep{alarm}, with $37$ nodes, $46$ arcs and $p = 509$
    parameters. A BN designed by medical experts to provide an alarm message
    system for intensive care unit patients based on the output a number of
    vital signs monitoring devices.
  \item HEPAR II \citep{hepar2}, with $70$ nodes, $123$ arcs and $p = 1453$
    parameters. A BN model for the diagnosis of liver disorders from related
    clinical conditions (e.g., gallstones) and relevant biomarkers (e.g.,
    bilirubin, hepatocellular markers).
  \item ANDES \citep{andes}, with $223$ nodes, $338$ arcs and $p = 1157$
    parameters. An intelligent tutoring system based on a student model for
    the field of classical Newtonian physics, developed at the University of
    Pittsburgh and at the United States Naval Academy. It handles long-term
    knowledge assessment, plan recognition, and prediction of students' actions
    during problem solving exercises.
  \item LINK \citep{link}, with $724$ nodes, $1125$ arcs and $p = 14211$
    parameters. Developed in the context of linkage analysis in large pedigrees,
    it models the linkage and the distance between a causal gene and a genetic
    marker.
  \item MUNIN \citep{munin}, with $1041$ nodes, $1397$ arcs with $p = 80592$
    parameters. A BN designed by experts to interpret results from 
    electromyographic examinations and diagnose a large set of common diseases
    from physical conditions such as atrophy and active nerve fibres.
\end{itemize}
Simulations were performed on a cluster of $7$ dual AMD Opteron 6136, each
with $16$ cores and $78$GB of RAM, with \proglang{R} 3.1.0 and \pkg{bnlearn}
3.5. For each BN, we considered $6$ different sample sizes ($n = 0.1p, 0.2p,
0.5p, p, 2p, 5p$), chosen as multiples of $p$ to facilitate comparisons between
networks of such different complexity; and $4$ different constraint-based
structure learning algorithms (GS, Inter-IAMB, MMPC, SI-HITON-PC). Since all 
reference BNs are discrete, we used the asymptotic $\chi^2$ mutual information
test with $\alpha = 0.01$. For each combination of BN, sample size and
algorithm we repeated the following simulation $20$ times. First, we loaded the
BN from the RDA file downloaded from the repository (\url{alarm.rda} below)
and generated a sample of the appropriate size with \code{rbn}.
\begin{Code}
R> load("alarm.rda")
R> sim = rbn(bn, n = round(0.1 * nparams(bn)))
\end{Code}
From that sample, we learned the skeleton of the DAG with 
(\code{optimized = TRUE}) and without backtracking (\code{optimized = FALSE}).
\begin{Code}
R> skel.orig = skeleton(si.hiton.pc(sim, test = "mi", alpha = 0.01,
+                         optimized = FALSE))
R> skel.back = skeleton(si.hiton.pc(sim, test = "mi", alpha = 0.01,
+                         optimized = TRUE))
\end{Code}

\begin{figure}[p]
  \begin{center}
  \includegraphics[width=\textwidth]{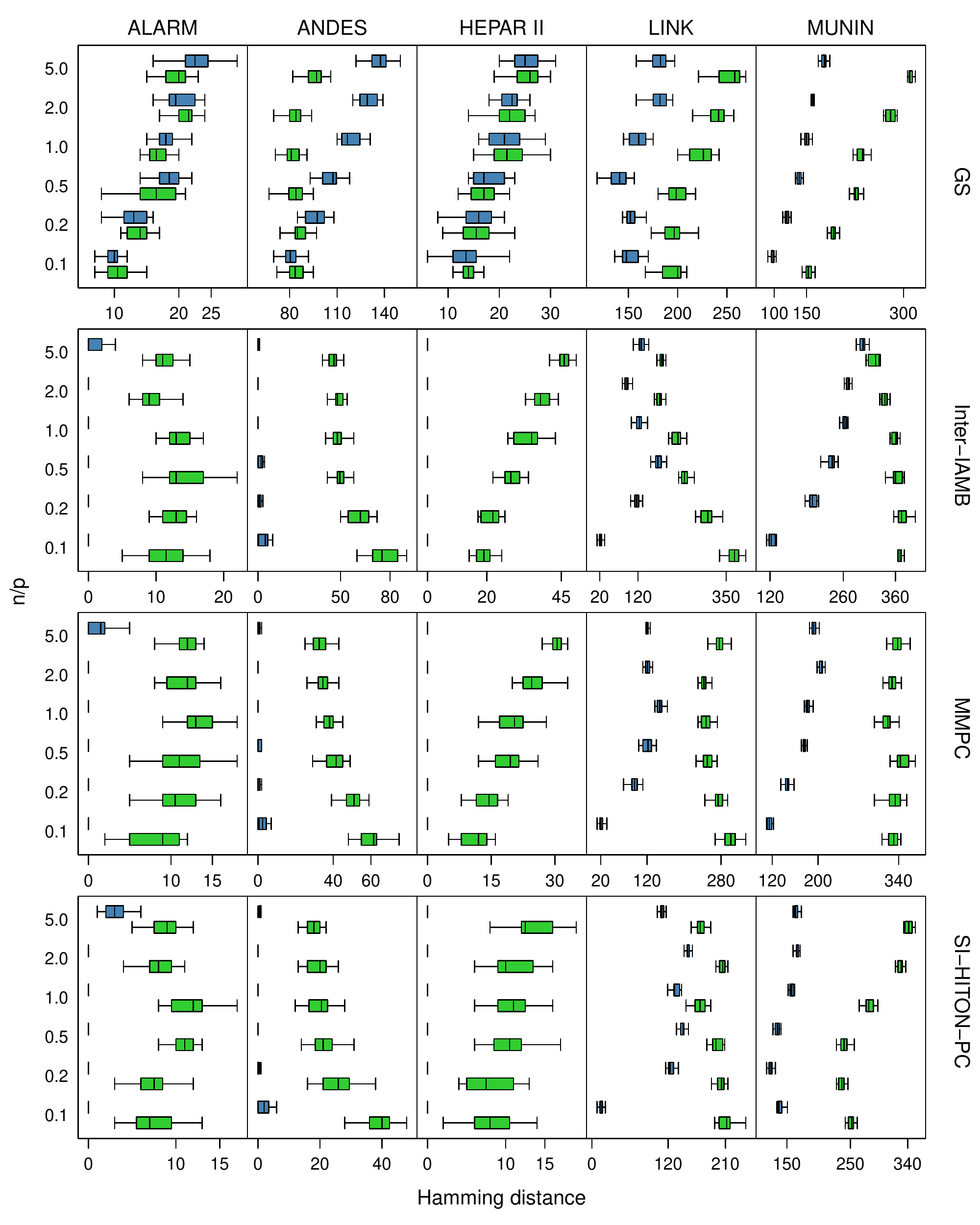}
  \caption{Hamming distance between skeletons learned for the ALARM, ANDES,
    HEPAR II, LINK and MUNIN reference BNs before and after reversing the
    ordering of the variables, for various $n/p$ ratios and algorithms. Blue
    boxplots correspond to structure learning without backtracking, green
    boxplots to learning with backtracking.}
  \label{fig:flip}
  \end{center}
\end{figure}

Subsequently, we reversed the order of the columns in the data to investigate
whether this results in a different skeleton.
\begin{Code}
R> revsim = sim[, rev(seq(ncol(sim)))]
\end{Code}

After learning the skeleton with (\code{rskel.back}) and without backtracking
(\code{rskel.orig}) from \code{revsim}, we compared the output with that from
\code{sim} using Hamming distance \citep{graphs}.
\begin{Code}
R> hamming(skel.orig, rskel.orig)
[1] 0
R> hamming(skel.back, rskel.back)
[1] 10
\end{Code}
Ideally, \code{skel.orig} and \code{rskel.orig} should be identical and
therefore their Hamming distance should be zero. This may not be the case for
BNs with deterministic 0-1 nodes, whose structure is unlikely to be learned
correctly by any of the considered algorithms; or when conditional independence
tests are biased and have low power because of small sample sizes. The
difference between the Hamming distance of \code{skel.orig} and 
\code{rskel.orig} and that of \code{skel.back} and \code{rskel.back} gives an
indication of the dependence on the ordering of the variables introduced by
backtracking. It is important to note that different algorithms will also learn
the structure of the reference BNs with varying degrees of accuracy, as 
described in the original papers and in \citet{hitonpc}. However, in this paper
we choose to focus on the effect of backtracking (and later of parallelisation)
as an algorithm-independent optimisation technique. Therefore, we compare
\code{skel.orig} with \code{rskel.orig} and \code{skel.back} with 
\code{rskel.back} instead of comparing all of them to the true skeleton of each
reference BN.

The results of this simulation are shown in Figure \ref{fig:flip}. With the
exception of ALARM, ANDES, HEPAR II for the GS algorithm, the Hamming distance
between the learned BNs is always greater when backtracking is used. In other
words, \code{hamming(skel.back, rskel.back)} is greater than 
\code{hamming(skel.orig, rskel.orig)} for all BNs, algorithms and sample sizes.
In fact, Hamming distance does not appear to converge to zero as sample size
increases; on the contrary, large samples make even weak dependencies 
detectable and thus increase the chances of getting different skeletons. This
trend is consistently more marked when using backtracking, as is the range of
observed Hamming distances in each configuration of BN, sample size and 
learning algorithm. The combination of this two facts suggests that backtracking
does indeed make learning dependent on the order in which the variables are
considered; and that it increases the variability of the learned structure.

\section{A framework for parallel constraint-based learning}
\label{sec:parallel}

\begin{figure}[t]
  \begin{center}
  \includegraphics[width=\textwidth]{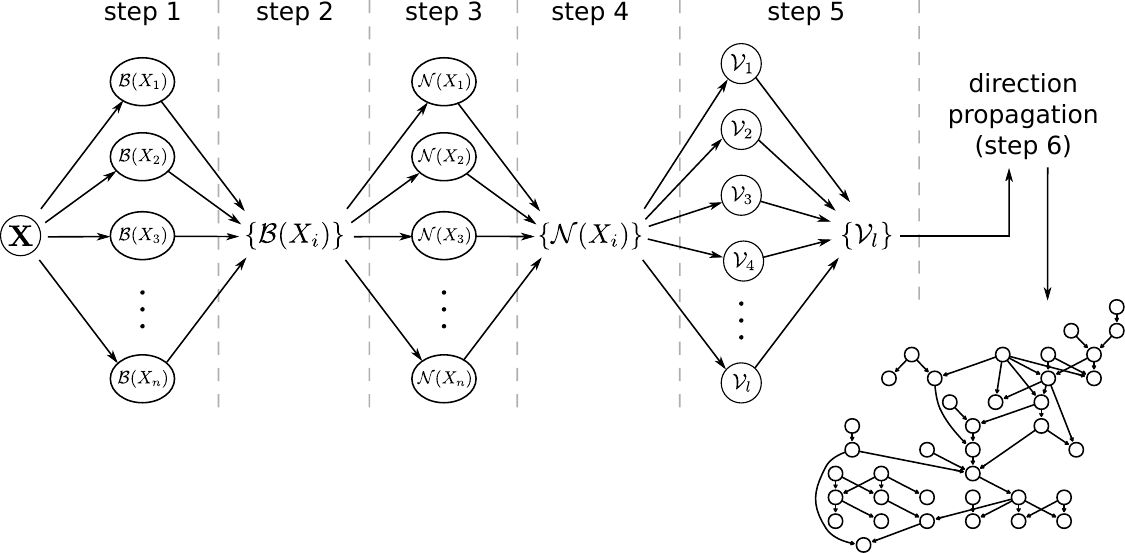}
  \caption{Software architecture for parallel constraint-based structure
    learning; parallel implementation of Algorithm \ref{algo:ic} in
    \pkg{bnlearn}.}
  \label{fig:parallel}
  \end{center}
\end{figure}

Constraint-based algorithms as described in Algorithm \ref{algo:ic} display
\textit{coarse-grained parallelism}: different parts need to be synchronised
only three times, in steps \ref{pt2}, \ref{pt4} and \ref{pt6}. Steps \ref{pt1},
\ref{pt3} and \ref{pt5} are \textit{embarrassingly parallel}, because each
$\B(X_i)$, each $\N(X_i)$ and each $\V_l$ can be learned independently from
the others. In practice, this means changing step \ref{pt1} from 
\begin{Code}
R> sapply(names(learning.test),
+           function(node) {
+             learn.mb(learning.test, node = node, method = "si.hiton.pc")
+           })
\end{Code}
to 
\begin{Code}
R> library(parallel)
R> cl = makeCluster(2)
R> clusterEvalQ(cl, library(bnlearn))
R> parSapply(cl, names(learning.test),
+          function(node) {
+            learn.mb(learning.test, node = node, method = "si.hiton.pc")
+          })
\end{Code}
using the functionality provided by the \pkg{parallel} package \citep{rcore}.
Step \ref{pt3} can be modified in the same way, just calling \code{learn.nbr}
instead of \code{learn.mb}. Step \ref{pt6} on the other hand is 
\textit{inherently sequential} because of its iterative formulation.
Parallelising Algorithm \ref{pt1} on a step-by-step basis is therefore very
convenient. As shown in Figure \ref{fig:parallel}, the implementation still
follows the same steps; it performs exactly the same conditional independence
tests, thus resulting in the same BN; and it can scale efficiently because
computationally intensive steps can be partitioned in as many as parts as
there are variables. Splitting the tests in large batches corresponding to
the $\B(X_i)$, $\N(X_i)$ and $\V_l$ also reduces the amount of information
exchanged by different parts of the implementation, reducing overhead.  

Similar changes could in principle be applied to the PC algorithm; different
pairs of nodes can be analysed in parallel and arcs merged into the $\N(X_i)$
at the end of step \ref{pt3}. However, as was the case for backtracking, the
monolithic implementation in \pkg{pcalg} would require a complete refactoring
beforehand.

\subsection{Simulations on the reference BNs}
\label{sec:reference}

All constraint-based learning algorithms in \pkg{bnlearn} have such a parallel
implementation made available transparently to the user, who only needs to
initialise a \code{cluster} object using \pkg{parallel}. The master \proglang{R}
process controls the learning process and distributes the $\B(X_i)$, $N(X_i)$
and $\V_l$ to the slave processes, executing only steps \ref{pt2}, \ref{pt4}
and \ref{pt6} itself. Consider, for example, the Inter-IAMB algorithm and the
data set generated from the ALARM reference BN shipped with \pkg{bnlearn}. 
\begin{verbatim}
R> data("alarm")
R> library("parallel")
R> cl = makeCluster(2)
R> res = inter.iamb(alarm, cluster = cl)
R> unlist(clusterEvalQ(cl, test.counter()))
[1] 3637 3743
R> stopCluster(cl)
\end{verbatim}
After generating a cluster \code{cl} with $2$ slave processes with
\code{makeCluster}, we passed it to \code{inter.iamb} via the \code{cluster}
argument. As we can see from the output of \texttt{clusterEvalQ}, the first
slave process performed $3637$ ($49.3\%$) conditional tests, and the second
$3743$ ($50.7\%$). The difference in the number of tests between the two
slaves is due to the topology of the BN: the $\B(X_i)$ and $\N(X_i)$ have
different sizes and therefore require different numbers of tests to learn.
This in turn also affects the number of tests required to learn the
v-structures $\V_l$.

\begin{figure}[p]
  \begin{center}
  \includegraphics[width=0.8\textwidth]{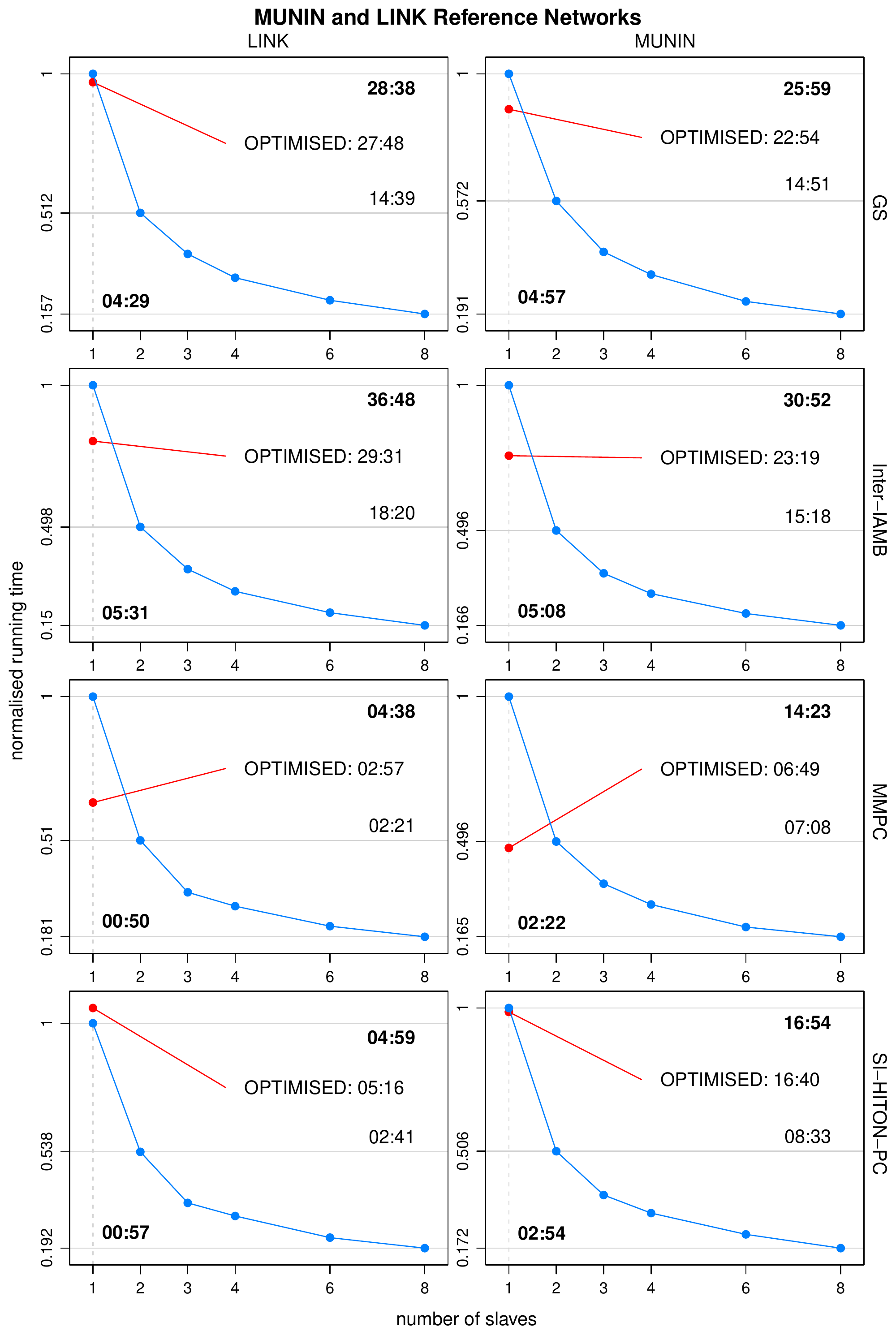}
  \caption{Normalised running times for learning the skeletons of the MUNIN
    and LINK reference BNs with the GS, Inter-IAMB, MMPC and SI-HITON-PC
    algorithms. Raw running times are reported for backtracking and for
    parallel learning with $1$, $2$ and $8$ slave processes.}
  \label{fig:benchmark}
  \end{center}
\end{figure}

\begin{figure}[p]
  \begin{center}
  \includegraphics[width=0.8\textwidth]{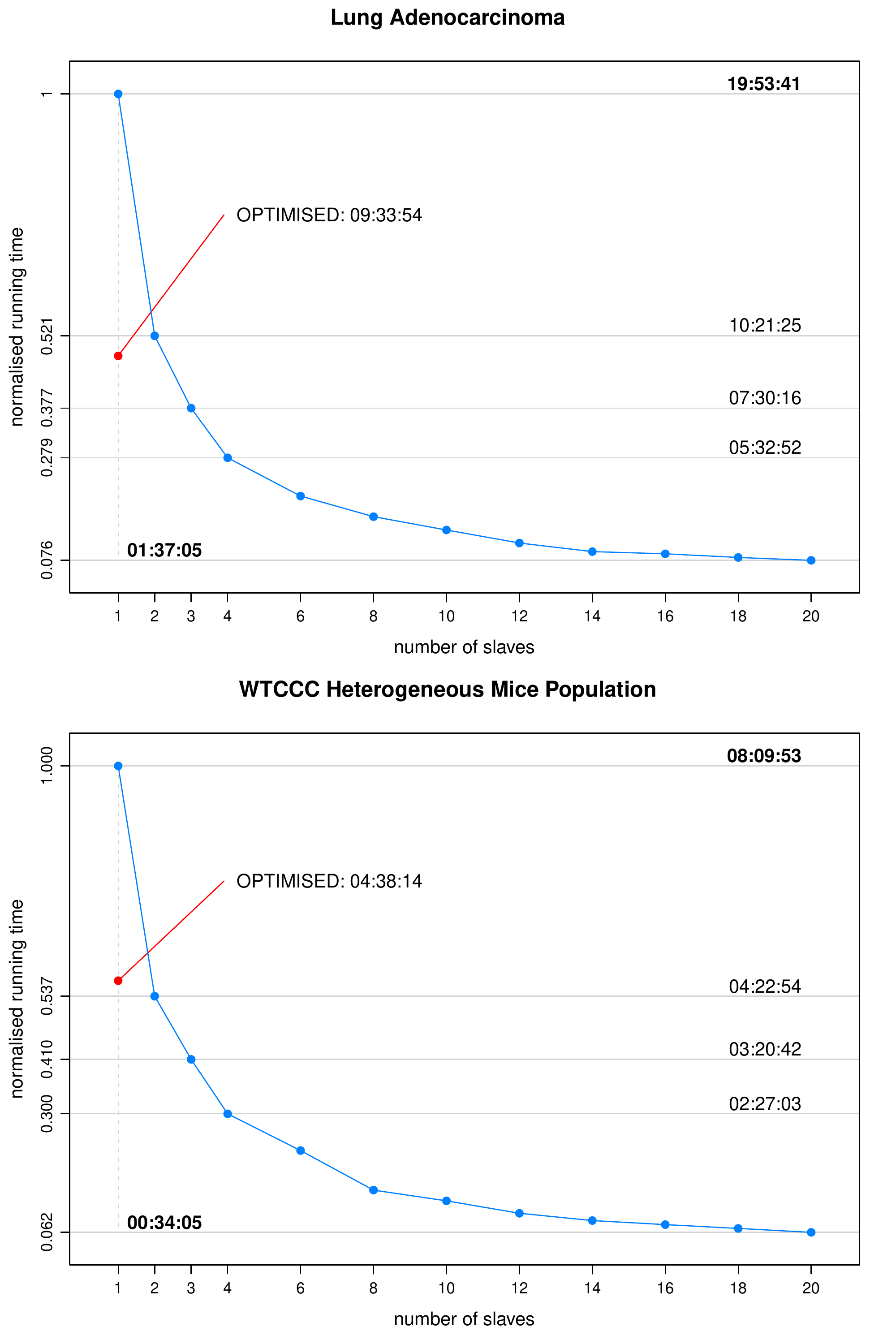}
  \caption{Running times for learning the skeletons underlying the lung
    adenocarcinoma \citep{adenocarcinoma} and mice \citep{mice2} data sets
    with SI-HITON-PC. Raw running times are reported for backtracking and
    for parallel learning with $1$, $2$, $3$, $4$ and $20$ slave processes.}
  \label{fig:realworld}
  \end{center}
\end{figure}

Increasing the number of slave processes reduces the number of tests performed
by each of them, further increasing the overall performance of the algorithm.
\begin{verbatim}
R> cl = makeCluster(3)
R> res = inter.iamb(alarm, cluster = cl)
R> unlist(clusterEvalQ(cl, test.counter()))
[1] 2218 2479 2683
R> stopCluster(cl)
R> cl = makeCluster(4)
R> res = inter.iamb(alarm, cluster = cl)
R> unlist(clusterEvalQ(cl, test.counter()))
[1] 1737 1900 1719 2024
R> stopCluster(cl)
\end{verbatim}
The raw and normalised running times of the algorithms used in Section
\ref{sec:backtracking} are reported in Figure \ref{fig:benchmark} for clusters
of $1$, $2$, $3$, $4$, $6$ and $8$ slaves; values are averaged over $10$ runs
for each configuration using generated data sets of size $20000$. Only the
results for LINK and MUNIN are shown, as they are the largest reference BNs
considered in this paper. For ALARM, HEPAR II and ANDES, and for smaller sample
sizes, running times are too short to make parallel learning meaningful for at
least MMPC and SI-HITON-PC. It is clear from the figure that the gains in running
time follow the \emph{law of diminishing returns}: adding more slaves produces
smaller and smaller improvements. Furthermore, tests are never split uniformly
among the slave processes and therefore slaves that have fewer tests to perform
are left waiting for others to complete (see, for instance, the \proglang{R}
code snippets above). Even so, the parallel implementations in \pkg{bnlearn}
scale efficiently up to $8$ slaves. In the absence of any overhead we would
expect the average normalised running time to be approximately $1/8 = 0.125$;
observed values are in the range $[0.157, 0.191]$. The difference, which is in
the range $[0.032, 0.066]$, can be attributed to a combination of communication
and synchronisation costs as discussed above. Optimised learning is at best 
competitive with $2$ slaves (MMPC, MUNIN), and at worst may actually degrade
performance (LINK, SI-HITON-PC).

\subsection{Simulations on the real-world data}
\label{sec:real}

To provide a more realistic benchmarking on large-scale biological data, 
we applied SI-HITON-PC on the lung adenocarcinoma gene expression data ($86$
observations, $7131$ variables representing expression levels) from 
\cite{adenocarcinoma}; and on the Wellcome Trust Case Control Consortium 
(WTCCC) heterogeneous mice sequence data ($1940$ observations, $4053$ variables
representing allele counts) from \cite{mice2}. The former is a landmark study
in predicting patient survival after an early-stage lung adenocarcinoma
diagnosis. Building a gene network from such data to explore interactions and
the presence of regulator genes is a common task in systems biology literature,
hence the interest in benchmarking BN structure learning. The latter is a 
reference data set produced by WTCCC to study genome-wide high-resolution
mapping of quantitative trait loci using mice as animal models for human 
diseases. In this context BNs have been used to investigate dependence
patterns between single nucleotide polymorphisms \citep{morota}.

Both data sets are publicly available and have been preprocessed to remove
highly correlated variables ($\COR > 0.95$) and to impute missing values with
the \pkg{impute} package \citep{impute}. The adenocarcinoma data set has a
sample size which is extremely small compared to the number of variables,
which is common in systems biology. On the other hand, the mice data has a
sample size that is typical of large genome-wide association studies. We ran
SI-HITON-PC using $1$, $2$, $3$, $4$, $6$, $8$, $10$, $12$, $14$, $16$, $18$
and $20$ slaves, averaging over $5$ runs in each case; the other algorithms
we considered in Section \ref{sec:backtracking} did not scale well enough to
handle either data set. Variables were treated as continuous, and independence
was tested using the Student's $t$ test for correlation with $\alpha = 0.01$.

As we can see in Figure \ref{fig:realworld}, we observe a low overhead even for
$20$ slave processes, with normalised running times of $0.062$ (mice) and $0.076$
(adenocarcinoma) which are very close to the theoretical $1/20 = 0.05$. Similar
considerations can be made across the whole range of $2$ to $20$ slaves, with a
measured overhead between $0.02$ and $0.08$. Surprisingly, overhead seems to 
decrease in absolute terms with the number of slaves, from $0.04$ (adenocarcinoma)
and $0.08$ (mice) for $3$ slaves to $0.012$ (adenocarcinoma) and $0.026$ (mice)
for $20$ slaves. However, clusters with  $2$ slaves have a smaller overhead
($0.021$ and $0.037$) than those with $3$ or $4$ slaves. We note that overhead
is comparable to that of the reference BNs in Section \ref{sec:backtracking},
suggesting that it does not strongly depend on the size of the BN; and that the
widely different sample sizes of the two data sets also seem to have little
effect. Again, the running time of the optimised implementation is comparable
with that of $2$ slaves.

\section{Discussion and conclusions}

In this paper we described a software architecture and framework to create
parallel implementations of constraint-based structure learning algorithms for
BNs and its implementation in \pkg{bnlearn}. Since all these algorithms trace
their roots to the IC algorithm from \citet{ic}, they share a common layout
and can be parallelised in the same way. In particular, several steps are
embarrassingly parallel and can be trivially split in independent parts to be
executed simultaneously. This is important for two reasons. Firstly, it limits
the amount of overhead in the parallel computations due to the need of keeping
different parts of the algorithms in sync. As we have seen in Section
\ref{sec:parallel}, this allows the parallel implementations to scale
efficiently in the number of slave processes. Secondly, it implies that the
parallel implementation of each algorithm performs the same conditional
independence tests as the original. This is in contrast with backtracking,
which is the only widespread way of improving the sequential performance of
constraint-based algorithms. Different approaches to backtracking have different
speed-quality tradeoffs, which motivated the adoption of that currently
implemented in \pkg{bnlearn}. The simulations in Section \ref{sec:backtracking}
suggest that backtracking can increase the variability of the DAGs learned from
the data. At the same time, speed gains are competitive at most with the parallel
implementation with $2$ slave processes. Since most computers in recent times
come with at least $2$ cores, it is possible to outperform backtracking even on
commodity hardware while retaining the lower variability of the non-optimised
implementations. Furthermore, even for the largest number of processes
considered in this paper ($8$ for the reference BNs, $20$ for the real-world
data), the overhead introduced by communication and synchronisation between the
slaves is low; the highest observed value is $0.08$. This suggests that the
proposed software architecture as implemented in \pkg{bnlearn} and \pkg{parallel}
scales efficiently for the range of sample sizes and number variables considered
in Section \ref{sec:parallel}. Finally, it is important to note that these
considerations arise from both discrete and Gaussian BNs and a variety of
constraint-based structure learning algorithms.

As for future research, there are several possible ways in which the current
implementation may be studied and improved. First of all, overhead might be
reduced by replacing \code{parSapply} with a function that allocates the
$\B(X_i)$ and $\N(X_i)$ dynamically to the slaves as they become idle. Assuming
the underlying BN is sparse, which is often formalised with a bound on the size
of the $\B(X_i)$, this is likely to provide little practical benefit as the overhead
is already low compared to the gains provided by parallelism. However, there are
some specific settings such as gene regulatory networks \citep[e.g.,][]{regulators}
in which this assumption is known not to hold; improvements may then be 
substantial. Such a setup could be based either on the \code{mcparallel} and 
\code{mccollect} functions in the \pkg{parallel} package, which unfortunately
are not available on Windows, or by avoiding \pkg{parallel} entirely to use the
\pkg{Rmpi} package directly \citep{rmpi}. Synchronisation in steps \ref{pt2} and
\ref{pt4} is required to obtain a consistent BN and thus precludes the use of 
partial update techniques such as that described in \citet{lda}.

It would also be interesting to consider how the overhead scales in the sample
size and in the complexity of the BN. On average, the number of conditional
independence tests required by constraint-based algorithms scales quadratically
in the number of variables; and the tests themselves are typically linear in
complexity in the sample size. Increasing or decreasing the latter should have
little impact on the overhead of parallel learning, because data need to be
copied only once to the slaves and that copy could be avoided altogether by
using a shared-memory architecture. The results in Section \ref{sec:real} 
suggest this is indeed the case, and the worst-case overhead is also similar
to that of reference BNs in Section \ref{sec:reference}. No locking or 
synchronisation is needed since the data are never modified by the algorithms.
On the other hand, the number of variables in the BN can affect overhead in 
various ways. If the BN is small, differences in the learning times of the 
$\B(X_i)$ and $\N(X_i)$ are more likely to leave slave processes idle even
with the dynamic allocation scheme described above. However, if the BN is large
the size of the $\B(X_i)$ and $\N(X_i)$ may vary dramatically thus introducing
significant overhead. In both cases the number of variables can only be used a
rough proxy for the complexity of the BN, which depends mainly on its topology;
and imposing sparsity assumptions on the structure of the BN can be used as a
tool to control overhead by keeping the $\B(X_i)$ and $\N(X_i)$ small and of
comparable size.

\end{document}